\documentclass[10pt,journal,compsoc]{IEEEtran}
\ifCLASSOPTIONcompsoc
  \usepackage[nocompress]{cite}
\else
  \usepackage{cite}
\fi

\ifCLASSINFOpdf
\else
\fi

\usepackage[vlined,ruled]{algorithm2e}
\usepackage{slashbox}
\usepackage{graphicx}
\usepackage{epstopdf}
\usepackage{subfigure}
\usepackage{amsmath}
\usepackage{amssymb}
\usepackage{url}
\usepackage{alltt}
\usepackage{arydshln}
\usepackage{listings}
\usepackage[usenames,dvipsnames]{color}
\usepackage{tikz}
\usepackage{mathtools}
\usetikzlibrary{arrows}
\newcommand{\ie}{{\em i.e.}}
\newcommand{\eg}{{\em e.g.}}
\newcommand{\etal}{{\em et.al.}}
\newcommand{\etc}{{\em etc.}}

\begin{document}

\title{Probabilistic Modeling and Inference for Obfuscated Cyber Attack Sequences}

\author{Haitao~Du,~\IEEEmembership{Member,~IEEE}
        and~Shanchieh~Jay~Yang,~\IEEEmembership{Senior~Member,~IEEE,}
\IEEEcompsocitemizethanks{\IEEEcompsocthanksitem Department of Computer Engineering, 
Rochester Institute of Technology, Rochester, New York 14623.\protect\\
E-mail: jay.yang@rit.edu
}
}

\markboth{IEEE Transactions on Emerging Topics in Computing,~Special Issue}%
{Yang \MakeLowercase{\textit{et al.}}: Probabilistic Modeling and Inference for Obfuscated Cyber Attack Sequences}

\IEEEtitleabstractindextext{%
\begin{abstract}
A key element in defending computer networks is to recognize the types of cyber attacks based on the observed malicious activities. Obfuscation onto what could have been observed of an attack sequence may lead to mis-interpretation of its effect and intent, leading to ineffective defense or recovery deployments. This work develops probabilistic graphical models to generalize a few obfuscation techniques and to enable analyses of the Expected Classification Accuracy (ECA) as a result of these different obfuscation on various attack models. Determining the ECA is a NP-Hard problem due to the combinatorial number of possibilities. This paper presents several polynomial-time algorithms to find the theoretically bounded approximation of ECA under different attack obfuscation models. Comprehensive simulation shows the impact on ECA due to alteration, insertion and removal of attack action sequence, with increasing observation length, level of obfuscation and model complexity. 
\end{abstract}

}

\maketitle
\IEEEdisplaynontitleabstractindextext
\IEEEpeerreviewmaketitle
\IEEEraisesectionheading{\section{Introduction}\label{sec:introduction}}
\IEEEPARstart{T}{he} increasing vulnerabilities and freely distributed cyber attack tools have led to significant volume of malicious activities from the Internet to penetrate enterprise networks. Ptacek \cite{Ptacek1998} pointed out that attack obfuscation is inevitable because the inherent limitations of Network Intrusion Detection Systems (NIDS) which monitor and match the target system response. It is impossible to represent the exact responses of the ever-increasing operating systems, applications, and communication protocols, so some malicious actions will not be observed. The signature-based nature of common NIDS requires pattern matching of observed actions to known malicious signatures, and, hence, is susceptible to attack obfuscation that can present alternative pattern. The uses of source IP spoofing \cite{Lyon2009a}, which hides the real identity of the attacker, and compromised machines as stepping stones \cite{Fuchsberger2005}, allows the attacker to either hide crucial actions and/or inject irrelevant actions to distract analysts with a large number of actions from various origins.

The possibilities to remove, insert, and alter observables of malicious actions present significant challenges for cyber defense that aims at analyzing and determining the effect and intent of a cyber attack based on its manifestation. Research works beyond intrusion detection have presented novel methods to extract and correlate observables of cyber attacks to determine their behavior or impact. Examples in this area include those utilizing Dynamic Bayesian Networks \cite{Du2014}, Variable Length Markov Models \cite{Du2010}, Attack Graphs \cite{Wang2006}, and Attack Social Graphs \cite{Du2011f}. A detailed summary of how some of these methods can be used to project cyber attacks is given by Yang et al. \cite{Yang2014}. These methods might not work as well when facing aforementioned and other obfuscation techniques. It is unclear what level of effect each type of obfuscation will have on these attack modeling methods. This work, thus, aims at analyzing formally and generally how obfuscation affects the accuracy of attack modeling; evaluating which obfuscation techniques will have a higher impact; and discovering the factors may change their impact on attack modeling.

A general set of probabilistic graphical models is developed to represent how cyber attacks may transpire with and without obfuscation action removal, insertion, and alteration. These models enable formal analyses of the effect of obfuscation to correctly classify an observed sequence of attack actions to the attack model as if there were no obfuscation. A formal metric is defined to assess the Expected Classification Accuracy (ECA) based on the concept of Bayes error. In order to compute the expected value with exponentially large number of scenarios, this work develops efficient algorithms based on dynamic programming and Monte-Carlo sampling to perform approximate inference. Comprehensive simulation on various combinations of attack models and obfuscation techniques provides insights of how each obfuscation affects ECA under different observation sequence length, obfuscation level, and attack model complexity.

The rest of the paper is organized as follows. Section \ref{sec:obfuscation} gives a formal definition of attack models with and without the three obfuscation techniques. Section \ref{sec:probinf} presents the problem formulation of calculating ECA, and the efficient algorithms that find the theoretically bounded approximations of ECA given the attack and obfuscation models. Section \ref{sec:simulation} illustrates the design of experiment and the simulation results. Finally, Section \ref{sec:conclusion} summarizes and concludes the work presented in this paper.

\section{Obfuscation and Formulation}
\label{sec:obfuscation}
\subsection{Attack Obfuscation Techniques}
Going beyond manipulating individual events, this paper focuses on how the attackers can use basic obfuscation actions together to achieve attack strategy level deception. This work considers three categories of obfuscation strategies: \textit{action alteration}, \textit{action insertion} and \textit{action removal}. These general categories are based upon experiences in working with security experts during the DARPA cyber insider threat project \cite{DARPACINDER}. The term \textit{noise attack sequence} is a general term and can be used to describe an observed alert sequence with intentional (attacker's obfuscation) or unintentional noise (such as IDS sensor failure). This work only considers the case of attacker's intentional obfuscation. The term \textit{noise} and \textit{obfuscation} will be interchangeably used. The term \textit{clean sequence} will be used to represent the original attack on selected target without using obfuscation techniques.

\subsubsection{Action alteration}
For signature-based detection engine, modifying the payload and craft a signature for intended alerts can be easily done. The attacker can alter alerts to hide the true origination and attack characteristic. Furthermore, sometimes, to achieve the same reconnaissance or intrusion objective, many actions can be interchangeable to be played. Changing the order of attacking actions can create equivalent sequence which can make the whole sequence more versatile and avoid being detected by matching to the classical intrusion sequence pattern. For example, the attacker can change the \textit{time-to-live} filed in ICMP ping, to generate alerts that look like to be originated from other OS platform, while achieving the same reconnaissance goal. Such obfuscation can be misleading for some alert correlation systems and cause alert correlation failure.

\subsubsection{Action insertion}
Inserting overwhelming alerts can separate related attack actions to affect the analysis engine, \eg,  increasing miss-classification of attack strategy. Even more, overwhelming alerts can cause Denial-of-Services (DoS) on the analysis engine, because the capacity of all alert analysis engine are limited \cite{Haines2003, Ptacek1998}.

There are many ways to perform noise injection. The simplest way is writing a script to keep performing scanning or getting sensitive file actions from a target to trigger the corresponding detection rules. Although such activity will easily expose attacker's IP and can be easily blocked by system administrator, using such simple tricks to injecting alerts on compromised host can be effective to dilute the original attack traces. In addition, the \textit{self-throttling} technique can be used to replay the actions happened before and hide the most recent intrusion state, \eg, host discovering, service scanning, privilege escalation, etc. Finally, \textit{Activity splitting} is another type of noise injection. Being aware of certain patterns of probing that can be triggered by some detection engines, one can split a malicious signature into multiple steps, and fragment actions can look normal. For example, a long sequence of failed log-in attempts is indicative to a dictionary-based password brute-force attack. Activity splitting will make such action more stealthy.

\subsubsection{Action removal}
Action removal is the obfuscation technique where an attacker hides critical actions that are indicative of the intrusion state. For example, Idle scanning \cite{Lyon2009a} takes advantage of predictable TCP sequence number vulnerability and can be completely anonymous when probing target host's servers. On the other hand, by carefully choosing encoding or encryption schemes, an attacker's critical actions may not be observed with traditional NIDS alerts. 

\subsection{Obfuscated Attack Formulation}
\subsubsection{Attack model}
We consider an attack strategy/model as a probabilistic sequence model, \eg,  Markov model or Hidden Markov Model (HMM) \cite{rabiner1989}, to describe the different possible attack actions and capture the casual relationship of attack actions using transition probabilities. More specifically, an attack sequence is described as a \textbf{vector of random variables} and each observation is an instance/sample of the attack model. When obfuscated, the attack sequence is modeled by another vector of random variables, where an \textit{obfuscation model} represents the obfuscation techniques probabilistically. The joint distribution is the overall description for the attack sequence that contains possible obfuscated observations. Because the clean attack sequence and the obfuscated sequence are not independent, one needs to jointly treat the attack model and the obfuscation model for probability inference.

Let discrete random variable $\Omega \in \{0,1,2,3,\cdots\}$ represents the set of possible attack actions (examples of attack actions can be found in Fig. \ref{fig:actionspace}), the attack sequence is defined as a length-$N$ vector random variable $\mathbf X$, where random variable $X_k \in \Omega, k \in \{1,2,3, \cdots, N\}$ is defined as the $k^{th}$ observed action in the attack sequence $\mathbf X$. An attack model is a probabilistic sequence model to specify $P(\mathbf X)$, which is shown in \eqref{eq:finiteJoint} as a $L^{th}$ order Markov model.

\begin{align}
P(\mathbf{X})= P(X_1,\cdots,X_L) \prod_{k=1}^{N-L}P(X_{L+k}| X_{L+k-1}, \cdots, X_{k})
\label{eq:finiteJoint}
\end{align}
where $P(X_1,\cdots,X_L)$ represents the initial distribution of the $L^{th}$ order Markov model. The Markov property (given $L$ observations past and further are independent) enables the product form decomposition of $P(\mathbf X)$.

The attack model discussed here does not take the obfuscated observations into account and it represents the intended attack strategy of the attacker. The term \textit{clean attack sequence} is used to represent the sequences directly generated from this attack model. 

Let a random variable vector $\mathbf Y$ represents obfuscated attack sequence. $\mathbf Y$ depends on $\mathbf X$ and $P(\mathbf Y|\mathbf X)$ describes the obfuscation model. As discussed in Section \ref{sec:obfuscation}, other than action alteration, in general $\mathbf X$ and $\mathbf Y$ in different length. Therefore, we categorize obfuscation into two types, according to the length of $\mathbf X$ and $\mathbf Y$.

\begin{itemize}
\item Type-I obfuscation model: $|\mathbf X|=|\mathbf Y|$
\item Type-II obfuscation model: $|\mathbf X|\neq|\mathbf Y|$
\end{itemize}

\subsubsection{Type-I model for action alteration}
\label{sec:type1model}
For Type-I model, \ie, the clean and noise sequence have the same length. One way to model the dependencies between $\mathbf X$ and $\mathbf Y$ is to apply HMM as shown in \eqref{eq:noiseModelHMM}.

\begin{equation}
P(\mathbf Y | \mathbf X)  = \prod_{k=1}^N P(Y_k|X_k) 
\label{eq:noiseModelHMM}
\end{equation}

In HMM, observed event at time $i$ only directly depends on the corresponding hidden state. The emission probability $P(Y_k=y|X_k=x)$ can be described by a discrete function, $g(x,y)$. HMM is not a perfect model for attack obfuscation because there is no direct parameter to describe the amount of obfuscations. In addition, the connections in graphical notation are only from $X_k$ to $Y_k$, which is limited in modeling real-world attacks.A more general obfuscation model $P(\mathbf Y|\mathbf X)$ may consider additional  constraints on $\mathbf Y$. For example, one extension of \eqref{eq:noiseModelHMM} can be shown in \eqref{eq:noiseModel}.

\begin{equation}
  P(\mathbf Y | \mathbf X)  = {\frac{1}{{{N \choose M}}}} I(|\mathbf Y-\mathbf X|_H=M)\prod_{\substack{k=1\\k:X_k\neq Y_k}}^N P(Y_k|X_k)
\label{eq:noiseModel}
\end{equation}
where $I(\cdot)$ is the indicator function, $|\cdot|_H$ represents the Hamming distance between the two vectors. 

The added parameter $M$ can be interpreted as an estimate of the percent ($M$ out of $N$) of attack actions the attacker may change actions. It serves to complement $g(x,y)$, which describes the preference on which obfuscation action is more likely to be chosen. The detail of the noise model shown in $\eqref{eq:noiseModel}$ is described in \cite{Du2014}. On the other hand, we also extend the basic HMM structure to second or higher-order on $\mathbf X$ instead of just the first-order. Note that the previous works on attack sequence modeling \cite{Du2010} \cite{Du2011b} \cite{Du2013a} have shown higher-order models are needed. Our extension also allows assessing the impact of obfuscation as a function of the percentage of attack actions altered, which will be described in the simulation section.

\subsubsection{Type-II model for action insertion and action removal}

For Type-II model, $\mathbf X$ and $\mathbf Y$ have different lengths and how much obfuscation exists will be directly reflected in model structure. This work uses regularized structure as an example to show the inference design. The model structure can be easily changed, which will be discussed at the end of this subsection.

\begin{figure}[ht!]
\begin{center}
\subfigure[Graphical representation for action insertion]{
\begin{tikzpicture}
  [->,>=stealth',shorten >=1pt,auto,node distance=3cm,thick,every node/.style={circle,draw,minimum size=1.2cm}]
  \node (n1) at (1,8) {$X_1$};
  \node [fill=none,draw=none] (n3) at (4,8)  {$\cdots$};
  \node[font=\tiny] (n4) at (7,8)  {$X_{N/2}$};

  \node (n5) at (1,6)  {$Y_1$};
  \node (n6) at (2.5,6)  {$Y_2$};
  \node [fill=none,draw=none] (n9) at (4,6)  {$\cdots$};
  \node [fill=none,draw=none] (n10) at (5.5,6)  {$\cdots$};
  \node[font=\tiny] (n11) at (7,6)  {$Y_{N-1}$};
  \node (n12) at (8.5,6)  {$Y_{N}$};
  
\path [->](n1) edge (n3);
\path [->](n3) edge (n4);

\path [->](n5) edge (n6);
\path [->](n6) edge (n9);
\path [->](n9) edge (n10);
\path [->](n10) edge (n11);
\path [->](n11) edge (n12);

\path [->](n1) edge (n5);
\path [->](n1) edge (n6);
\path [->](n4) edge (n11);
\path [->](n4) edge (n12);
\end{tikzpicture}
\label{fig:ch3_insertion_model}
}
\subfigure[Graphical representation for action removal]{
\begin{tikzpicture}
  [->,>=stealth',shorten >=1pt,auto,node distance=3cm,thick,every node/.style={circle,draw,minimum size=1.2cm}]
  \node (n1) at (1,8) {$X_1$};
  \node (n2) at (2.5,8)  {$X_2$};
  \node[fill=none,draw=none] (n5) at (4,8)  {$\cdots$};
  \node[fill=none,draw=none] (n6) at (5.5,8)  {$\cdots$};
  \node[font=\tiny] (n7) at (7,8)  {$X_{N-1}$};
  \node (n8) at (8.5,8)  {$X_N$};

  \node (n9) at (1,6)  {$Y_1$};
  \node[fill=none,draw=none] (n11) at (4,6)  {$\cdots$};
  \node[font=\tiny] (n12) at (7,6)  {$Y_{N/2}$};

\path [->](n1) edge (n2);
\path [->](n2) edge (n5);
\path [->](n5) edge (n6);
\path [->](n6) edge (n7);
\path [->](n7) edge (n8);

\path [->](n9) edge (n11);
\path [->](n11) edge (n12);

\path [->](n1) edge (n9);
\path [->](n2) edge (n9);

\path [->](n7) edge (n12);
\path [->](n8) edge (n12);
\end{tikzpicture}
\label{fig:ch3_removal_model}
}
\end{center}
\caption{Graphical representation for action insertion and removal}
\end{figure}
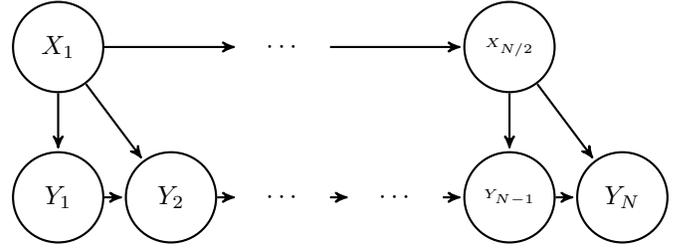
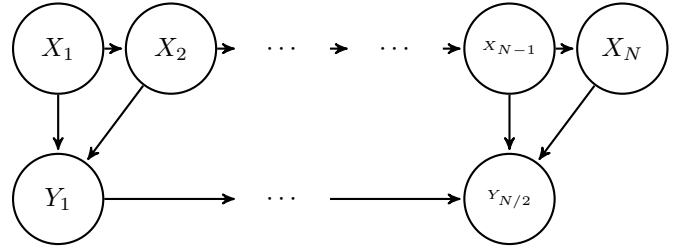

Figure \ref{fig:ch3_insertion_model} shows an example of the action insertion model, whereas for every clean action, one additional obfuscated action can be injected for a sequence $\mathbf Y$. In other words, the current action $Y_k$ conditionally depends on the previous action $Y_{k-1}$ and the clean attack action $X_{k/2}$.

The obfuscation model $\mathbf P(\mathbf Y| \mathbf X)$ for Fig. \ref{fig:ch3_insertion_model} can be described in \eqref{eqn:ch3_insertion_model}. Similarly, for the action removal case, the model structure for removing one out of every two clean actions is shown in Fig. \ref{fig:ch3_removal_model}. Its The obfuscation model is shown in \eqref{eqn:ch3_removal_model}.

\begin{align}
P(\mathbf Y|\mathbf X) & =P(Y_1|X_1)P(Y_2|Y_1,X_1) \notag\\
&\prod_{i=2}^{N/2} P(Y_{2i-1}|Y_{2i-2}, X_{i})P(Y_{2i}|Y_{2i-1}, X_{i})
\label{eqn:ch3_insertion_model}
\end{align}

\begin{equation}
P(\mathbf Y|\mathbf X)=P(Y_1|X_1, X_2)\prod_{i=2}^{N/2} P(Y_{i}|Y_{i-1}, X_{2i-1}, X_{2i})
\label{eqn:ch3_removal_model}
\end{equation}

The model structures shown here are examples to model cyber attack with obfuscations, but not intended to suggest that they are the most realistic or the only cases. This work focuses on how to treat models of similar structure to determine the limit to inference the clean actions from obfuscated ones.

For example, in the attack action alteration case (Type-I), one could model the clean sequence and noise sequence model as auto-regressive HMM \cite{murphy2002}. In such a case, noise action $Y_i$ depends on the previous noise action $Y_{i-1}$.  For the action insertion case (Type-II), one could model that the injected noise is conditionally independent of any variable, \ie, the attack randomly injects noise and there are no links between the injected noise and the true observations. Other examples for the Type-II model include extending it with a higher-order dependency on $\mathbf X$,  adding constrains on $\mathbf Y$, and removing dependencies by setting a special parameter of the general model.

\section{Probabilistic Inference for Obfuscated Attacks}
\label{sec:probinf}
\subsection{Definition: Expected Classification Accuracy (ECA)}
\label{sec:sec3problem_statement}
The attack obfuscation models proposed enable one to assess the impact under different attack obfuscation techniques. We first discuss about matching a clean attack sequence $\mathbf X$ to pre-defined models. The problem can be described as finding the attack model $C$ that maximize the posterior $P(C| \mathbf X)$. 

\begin{align}
\arg \max_C P(C| \mathbf X) &=\arg \max_C \frac {P(\mathbf X |C)P(C)} {P(\mathbf X)}\notag\\
&=\arg \max_C P(\mathbf X |C)P(C)
\label{eq:classifier}
\end{align}

To evaluate the impact caused by attack obfuscations, it is necessary to understand the performance limit of the classification when attacks contain obfuscation. This work proposes to use Expected Classification Accuracy (ECA) to assess how having obfuscated observations may affect the classification accuracy. ECA is closely related to many concepts in statistics, such as the \textit{irreducible error} or \textit{Bayes error rate} \cite{hastie2009}. This metric assumes that the \textit{true distribution} of attacks is known and expressed in statistical graph models. However, even with the model of the attack distribution, there may still be errors when making classification. This is because the true distribution of the different classes may \textit{overlap}. The term \textit{irreducible error} or \textit{Bayes error rate} describes how much the overlap is, and the ECA is defined as one minus Bayes error. Given an obfuscated sequence $\mathbf Y$, ECA is defined as

\begin{equation}
\text{ECA} \coloneqq \sum_\mathbf Y P(\mathbf Y)\max_C P(C|\mathbf Y)
\label{eq:optimalrateY}
\end{equation}

Equation \eqref{eq:optimalrateY} can be explained as follows: for any given obfuscated observation $\mathbf Y$, $P(C|Y)$ can be calculated for all attack models, and the noise sequence $\mathbf Y$ can be classified into pre-defined models by using $\arg \max_C P(C|\mathbf Y)$. By doing such classification, $\max_C P(C|\mathbf Y)$ percent of all classification results will be correct. Summing over all possible $\mathbf Y$ will give us the mathematical expectation of the classification accuracy.

Distributing $P(\mathbf Y)$ into the max operation, \eqref{eq:optimalrateY} can be written as 
\begin{align}
&\sum_\mathbf Y \max_C P(C|\mathbf Y)P(\mathbf Y) \notag\\
=&\sum_\mathbf Y \max_C P(C,\mathbf Y) =\sum_\mathbf Y \max_C P(\mathbf Y|C)P(C)
\label{eq:optimalrate2}
\end{align}

Here, the probability of a clean observation sequence for a given attack model $P(\mathbf X|C)$, the prior of attack models $P(C)$ and the noise model $P(\mathbf Y| \mathbf X)$ are assumed to be known. To calculate $\sum_\mathbf Y \max_C P(\mathbf Y|C)P(C)$, two sub-problems need to be solved:

\begin{itemize}
\item Given $P(\mathbf X|C)$ and $P(\mathbf Y|\mathbf X)$, calculate $P(\mathbf Y|C)$.
\item Given $P(\mathbf Y|C)$, calculate $\sum_\mathbf Y \max_C P(\mathbf Y|C)P(C)$.
\end{itemize}

The first subproblem can be solved by an extension of Message Passing Algorithm \cite{Bishop2006} using dynamic programming, and the second subproblem can be approximated by Monte-Carlo sampling with any desired precision and confidence. 

In summary, the problem addressed in our framework can be described as how to calculate the performance limit in the presence of obfuscated attack observations for different obfuscation strategy $P(\mathbf X|\mathbf Y)$. That is, calculate $\sum_\mathbf Y P(\mathbf Y)\max_C P(C|\mathbf Y)$, given $P(\mathbf X|C)$, $P(C)$, and $P(\mathbf Y| \mathbf X)$. The next subsection will show that this calculation, \ie, is computational challenging and requires efficient algorithms. Using the proposed algorithms to assess ECA, security analysts will be able to assess and weigh the potential impact of attack obfuscation under different scenarios. 

\subsection{Dynamic programming for solving sub-problem 1}

Probabilistic inference is the problem of calculating specific marginal or conditional distributions for a given model. The inference is computationally challenging because brute-force calculation needs to account for the exponential number of terms. Performing exact inference for arbitrary model structure of $P(\mathbf X)$ is a NP-hard problem \cite{murphy2002}. However, for some special structure of $P(\mathbf X)$ there exists efficient algorithm.

In HMM literature, there are classical algorithms to perform exact inference. For example, the Viterbi algorithm \cite{rabiner1989} efficiently calculates the most probable path of a clean sequence by solving \eqref{eq:infproblem1}.
\begin{equation}
\arg \max _{\mathbf X} P(\mathbf X| \mathbf Y)
\label{eq:infproblem1}
\end{equation}
Likewise, one can efficiently calculate $P(\mathbf Y)$ for HMM, because $P(\mathbf Y| \mathbf X)$ is relatively simple for HMM. However, some of the existing algorithms cannot be directly applied to the Type-I and Type-II models proposed in this paper. For Type-I model, constraint $M$ on $\mathbf Y$ will affect the possible values of $\mathbf X$ and the $\arg\max$ operation will only apply to a subset of $\mathbf X$ that satisfies the constraint. For Type-II model, the length of $\mathbf X$ and $\mathbf Y$ are different.

Here we discuss two simplified cases to show the idea of solving the inference problem for our models. After discussing the simplified problem, the algorithm design for general model strictures and extensions are given.

\subsubsection{Inference on a chain structure}
Suppose we want to solve the optimization problem shown in \eqref{eqn:ch4_inference_chain}.
\begin{equation}
\max_{\mathbf X}\prod_{k=1}^{N-1}f(X_{k},X_{k+1})
\label{eqn:ch4_inference_chain}
\end{equation}

The relationship for all the variables with a chain structure, where $X_i$ only \textit{interact} with $X_{i-1}$ and $X_{i+1}$. This simplified problem is very similar to the first-order Markov model: according to the Markovian property, $X_{i-1}$ and $X_{i+1}$ are independent given $X_i$. Because the objective function has such a special chain structure, one can use dynamic programming techniques to solve the optimization problem. Define a function $F_i(a)$ as the cost of the best length-$i$ subsequence that ends with a symbol $a$.

\begin{align}
F_{i}(a)= & \max_{X_{1}\cdots X_{i}} \prod_{k=1}^{i-1}f(X_{k},X_{k+1}) \label{eqn:fia1}\\
&\text{s.t.} ~X_{i}=a \notag
\end{align}

With the chain structure, the only \textit{connections} between the subsequence $X_1$ to $X_i$ and the subsequence $X_i$ to $X_N$ is the variable $X_i$. The reason we set the constrain of the subsequence ends with a symbol $a$ is because we want to \textit{decouple the interactions between two subsequences}. Such constrain will allow us to solve the problem in a smaller scale, and derive recursion rules for extension, which leads to the use of dynamic programming. 

Let $F_i(a)$ be the cost of the best length-$i$ subsequence and end with a symbol $a$ as shown in  \eqref{eqn:fia1}.  Equation \eqref{eqn:recursion1} gives the relationship between $F_i(a)$ and $F_{i-1}(a)$, that can be used in Algorithm \ref{alg:ch3_inf_chain} to find the optimal solution for \eqref{eqn:ch4_inference_chain}.

\begin{equation}
F_{i}(a)=\max_{b}F_{i-1}(b)\cdot f(b,a)
\label{eqn:recursion1}
\end{equation}

Algorithm \ref{alg:ch3_inf_chain} has a complexity of $\Theta(N\cdot|\Omega|^2)$, where $N$ is the length of the sequence and $|\Omega|$ is the number of the possible values of the random variables. This is a significant improvement over the brute-force approach over the exponential search space with $\Theta(|\Omega|^N)$ complexity.

\begin{algorithm}
\SetAlgoLined
\KwIn{Given the sequence length $N$, function $f(x,y)$, $X_i \in \Omega$}
\KwOut{$\max_{\mathbf X}\prod_{k=1}^{N-1}f(X_{k},X_{k+1})$}

\For{$a \in \Omega$}{
	Initialize $F_{2}(a)=\displaystyle\max_{X_{1}}f(X_{1},a)$
}

\For{$i \in {3,4,\cdots,N}$}{
	\For{$a \in \Omega$}{
		$\displaystyle F_{i}(a)=\max_{b}F_{i-1}(b)\cdot f(b,a)$
}
}
\Return {$\displaystyle\max_a(F_N(a))$}
\caption{Inference on a chain structure}
\label{alg:ch3_inf_chain}
\end{algorithm}

\subsubsection{Inference on a chain structure with constraints}

As discussed earlier, Type-I model has the additional constrain on $\mathbf Y$. Here, we add the constraint to the chain structure to illustrate the algorithm design. The revised problem is shown in \eqref{eqn:ch6_simplified_problem2}. The major difference is that only a subset of $\mathbf X$ needs to be considered. The subset of $\mathbf X$ depends on $M$ and $\mathbf Y$, \ie, only $M$ number of elements in $\mathbf X$ are allowed to be different from a given vector $\mathbf Y$. 

\begin{align}
\sum_{\mathbf X: |\mathbf X-\mathbf Y|_H=M}\prod_{i=1}^{N-1} f(X_{i+1},X_{i})
\label{eqn:ch6_simplified_problem2}
\end{align}

Comparing to the solution without the additional constraint, one can add another dimension to the dynamic programming table to decompose the dependencies on the constraint. We define function $F_{i,j}(a)$ as

\begin{align}
F_{i,j}(a) & =\sum_{X_1,\cdots,X_i}\prod_{k=1}^{i-1}f(X_k,X_{k+1})\label{eqn:ch4_constrain_dp_fun}\\
& \text{s.t.~~~~~}  X_i=a \notag\\
& \text{~~~~~~~~~} |<X_1,\cdots X_i> - <Y_1,\cdots Y_i>|_H=j \notag
\end{align}

Let $F_{i,j}(a)$ be the sum for \eqref{eqn:ch6_simplified_problem2} over the subsequence $X_1,\cdots,X_i$. In addition, the subsequence $X_1,\cdots,X_i$ is different from $Y_1,\cdots,Y_i$ by $j$ elements. The relationship between $F_{i,j}(a)$, $F_{i-1,j}(a)$ and $F_{i-1,j-1}(a)$ can be described in \eqref{eqn:recursion2}.

\begin{equation}
F_{i,j}(a)=
\begin{cases}
\displaystyle\sum_{b}F_{i-1,j}(b)\cdot f(b,a) & \text{if } a=Y_i \\
\\
\displaystyle\sum_{b}F_{i-1,j-1}(b)\cdot f(b,a) & \text{if } a\neq Y_i \\
\end{cases}\label{eqn:recursion2}
\end{equation}

\subsubsection{Calculating the noise sequence distribution for Type-I and Type-II models}
Using the recursion rule shown in \eqref{eqn:recursion2}, the complexity of the algorithm is $\Theta (N\cdot M \cdot |\Omega|^{2})$, where $N$ is the length of the sequence and $M$ is the total number of elements that are different between $\mathbf X$ and $\mathbf Y$. The complexity can be intuitively explained as follows: the dimension of the dynamic programming table is $N\times M$. For every entry, it stores a function of $a$. The calculation for an entry requires searching over the $\Omega$ space. In addition, evaluating the function needs to sum over the variable $b$ as shown in \eqref{eqn:recursion2}, which needs another loop over $\Omega$.

After discussing inference on a chain structure and inference with the additional constraint, we give the algorithm of calculating obfuscated sequence distribution for Type-I and Type-II models. To simplify the notation for the following discussion, we drop the random variable $C$ in $P(\mathbf Y|C)$ to focus on a specific and given attack model and obfuscation model. 

For the action alteration case, let $f(x,y)=P(X_{i+1}=y|X_i=x)$, $g(x,y)= P(Y_{i}=y|X_{i}=x)$. The dynamic programming function $F_{i,j}(\mathbf A)$ is defined in \eqref{eq:dpfdef}, where the length of vector $\mathbf A$ is $L$ and is used to decouple the $L^{th}$-order Markov model and the recursion rules are shown in \eqref{eqn:ch6_typeI_rule}. 

For the action insertion case, let $f(x,y)=P(X_{i+1}=y|X_i=x)$, $g(x,y,z)= P(Y_{2i-1}=z|Y_{2i-2}=x, X_{i}=y)$, $\phi(x,y,z)= P(Y_{2i}=z|Y_{2i-1}=x, X_{i}=y)$. The dynamic programming function is defined in \eqref{eqn:ch4_type2_dp_fun}, which gives the recursion rule.

For the action removal case, let $f(x,y)=P(X_{i+1}=y|X_i=x)$, $g(x,y,z,p)= P(Y_{i}=p|Y_{i-1}=x, X_{2i-1}=y, X_{2i}=z)$. Then we expand and re-write the product operation. The goal is to make the product subscript the same, so we can define the dynamic programming function. The dynamic programming function is shown in \eqref{eqn:ch4_removal_dp_fun}, with the recursion rules shown in \eqref{eqn:ch4_removal_rule}.

\begin{figure*}
\centering
\begin{align}
F_{i,j}(\mathbf A)=  &\sum_{{\mathbf{X}}: X_1,X_2, \cdots, X_i} \biggl( P(X_1,X_2, \cdots,X_L)\prod_{k=1}^{i-L}f(X_k, \cdots, X_{k+L}) \prod_{\substack{k=1\\i:X_k\neq Y_k}}^i g(X_k,Y_k) \biggr)\notag\\
\text {s.t.} ~ & <X_{i-L+1}, \cdots, X_{i}>=\mathbf A \notag\\
& |<X_1,\cdots,X_i>-<Y_1,\cdots,Y_i>|_H=j \label{eq:dpfdef}
\end{align}
\begin{equation}
F_{i,j}(\mathbf A)=\begin{cases}
\displaystyle\sum_{B}F_{i-1,j}(<B, \mathbf A'>)\cdot f(<B,\mathbf A>), & \text{if } A_L=Y_i \\
\\
\displaystyle\sum_{B}F_{i-1,j-1}(<B, \mathbf A'>)\cdot f(<B,\mathbf A>) g(A_L,Y_i), & \text{if } A_L\neq Y_i \\
\end{cases}
\label{eqn:ch6_typeI_rule}
\end{equation}
\begin{align}
F_i(a) =& \sum_{X_1,\cdots,X_i}\prod_{k=1}^{i-1}P(X_1)f(X_k,X_{k+1})\notag\\
&\prod_{k=2}^{i} P(Y_1|X_1)P(Y_2|Y_1, X_1) g(Y_{2k-2},X_k,Y_{2k-1})\phi(Y_{2k-1},X_{k}, Y_{2k}) \notag\\
& \text{s.t.~~~~~~~~~~~}  X_i=a \label{eqn:ch4_type2_dp_fun}
\end{align}
\begin{align}
F_i(a)=\sum_b F_{i-1}(b)\cdot f(b,a) \cdot g(Y_{2i-2},a, Y_{2i-1}) \cdot \phi(Y_{2i-1},a,Y_{2i})
\label{eqn:ch4_insertion_recursion}
\end{align}
\begin{align}
F_i(a) & =P(X_1)P(X_2|X_1)P(Y_1|X_1, X_2)\label{eqn:ch4_removal_dp_fun}\\
&\sum_{X_1,\cdots,X_{2i}} \prod_{k=2}^{i}f(X_{2k-2},X_{2k-1})f(X_{2k-1},X_{2k})\prod_{k=2}^{i} g(Y_{k-1}, X_{2k-1}, X_{2k},Y_{k})\notag\\
& \text{s.t.~~~~~~~~~~~}  X_{2i}=a \notag
\end{align}
\begin{align}
F_i(a)=\sum_{b,c} F_{i-1}(b)\cdot f(b,c)\cdot f(c,a) \cdot g(Y_{i-1},c, a, Y_i)
\label{eqn:ch4_removal_rule}
\end{align}
\caption{Recursion Rules for Action Insertion and Action Removal Inference Algorithm}
\label{fig:sub_prob1_formulas}
\end{figure*}

\subsection{Approximate inference for solving sub-problem 2}
This section discusses the second sub-problem: how to compute the ECA for noise sequence $\mathbf Y$, \textit{i.e.}, how to evaluate \eqref{eq:optimalrate2}, where $P(\mathbf Y|C)$ can be calculated from solving sub-problem 1 and $P(C)$ is known. Similar to the problem of calculating $P(\mathbf Y|C)$ addressed before, calculating the ECA in \eqref{eq:optimalrate2} also need to sum over an exponential number of terms. However, unlike solving the problem of $P(\mathbf Y|C)$, to the best of our knowledge, there is no efficient algorithm to solve \eqref{eq:optimalrate2} efficiently.

The challenge to solve \eqref{eq:optimalrate2} is due to the following. There is no analytical expression for $P(\mathbf Y|C)$; instead, for any given $\mathbf Y$, we need to execute Algorithm 1 to get $P(\mathbf Y|C)$. Without the appropriate structure to decompose the problem, the dynamic programming concept does not apply any more. In addition, the max operation makes the problem more complicated and prohibits the decoupling of the problem with an optimal structure.

\begin{algorithm}
\SetAlgoLined
\KwIn{Given $P(\mathbf X|C)$, $P(C)$, $P(\mathbf X|\mathbf Y)$, sample size $n$}
\KwOut{ECA for sequence classification}
Set $\mathbf S=<>$

\For {$i\in \{1,2,\cdots, n\}$}{
	Sample $C_0$ from $P(C)$

	Given sample $C_0$, sample $\mathbf X_0$ from $P(\mathbf X | C_0)$

	Given sample $\mathbf X_0$, sample $\mathbf Y_0$ from $P(\mathbf Y | \mathbf X_0)$

	Set $\mathbf V=<>$

	\For {$c\in \{1,2,\cdots, |C|\}$}{

	 Given $\mathbf Y_0$, calculate $P(\mathbf Y_0|c)$ 

	 Append $\mathbf V$ with $P(\mathbf Y_0|c)P(c)$

	}
	 Append $\mathbf S$ with ${\max(\mathbf V)}/{\text{sum}(\mathbf V)}$
}
\Return {${\max(\mathbf S)}/{\text{sum}(\mathbf S)}$}
\caption{Algorithm to estimate the ECA}
\label{alg:approximation}
\end{algorithm}

We propose to approximate $$\sum_\mathbf Y \max_C P(\mathbf Y|C)P(C)$$ with arbitrary precision and confidence using Monte-Carlo method. This is feasible because the samples of $\mathbf Y$ can be effectively drawn and the sample average can be used to estimate the mean as described in \eqref{eq:approx}.

\begin{equation}
\sum_{\mathbf Y} P(\mathbf Y) \psi(\mathbf Y) \approx \frac 1 n \sum_{{k=1}}^n \psi(\mathbf Y_k)
\label{eq:approx}
\end{equation}
where $\mathbf Y_k$ are independent and identically distributed random variables sampled from $P(\mathbf Y)$, and $\psi(\mathbf Y)$ is defined as $\max_C P(C|\mathbf Y)$ and can be calculated from $P(\mathbf Y|C)$
and $P(C)$. That is
\begin{equation}
\psi(\mathbf Y)=\max_C \frac {P(\mathbf Y|C)P(C)} {P(\mathbf Y)}=\max_C P(\mathbf Y|C)P(C)
\label{eq:psiY}
\end{equation}

Base on Hoeffding's bound \cite{serfling1974}, we have 

\begin{equation}
P\left(\left|\frac1 n \sum_{k=1}^n \psi(\mathbf Y_k)-E[\psi(\mathbf Y)]\right|>\epsilon \right)\leq \delta
\label{eq:hoeffding}
\end{equation}
where $n$ represents the number of samples and $\delta \triangleq 2\exp(-2n \epsilon^2)$.

Applying \eqref{eq:hoeffding} with $n\geq\frac 1 {2 \epsilon^2} \log \frac 2 \delta$, with probability at least $1 -\delta$, the difference between the approximation and the true value is at most $\epsilon$. In the experiments shown in this paper, we use 30,000 samples, which is larger than $26,492$ needed for $\epsilon=0.01$ and $\delta=0.01$.  The approximation algorithm is given in Algorithm \ref{alg:approximation}.

\subsection{Approximate Inference Algorithm Efficiency}

We verify the proposed algorithm as follows: for a small observation length, it is possible to calculate the exact ECA by summing over all $\mathbf X$. We did the exact calculation for observation length from $5$ to $10$ and use the proposed approximation algorithm for comparison. The simulation results show the estimation is better than the theoretical bounds. The theoretical bounds tell us the error should be smaller than $0.01$ with probability $99\%$. The experiments of $10$ repetitions show $0.003$ max error for different observation lengths. This is not surprising since the Hoeffeding bound is usually a loose bound. 

\begin{figure}[!ht]
\centering
\includegraphics[width=0.85\columnwidth] {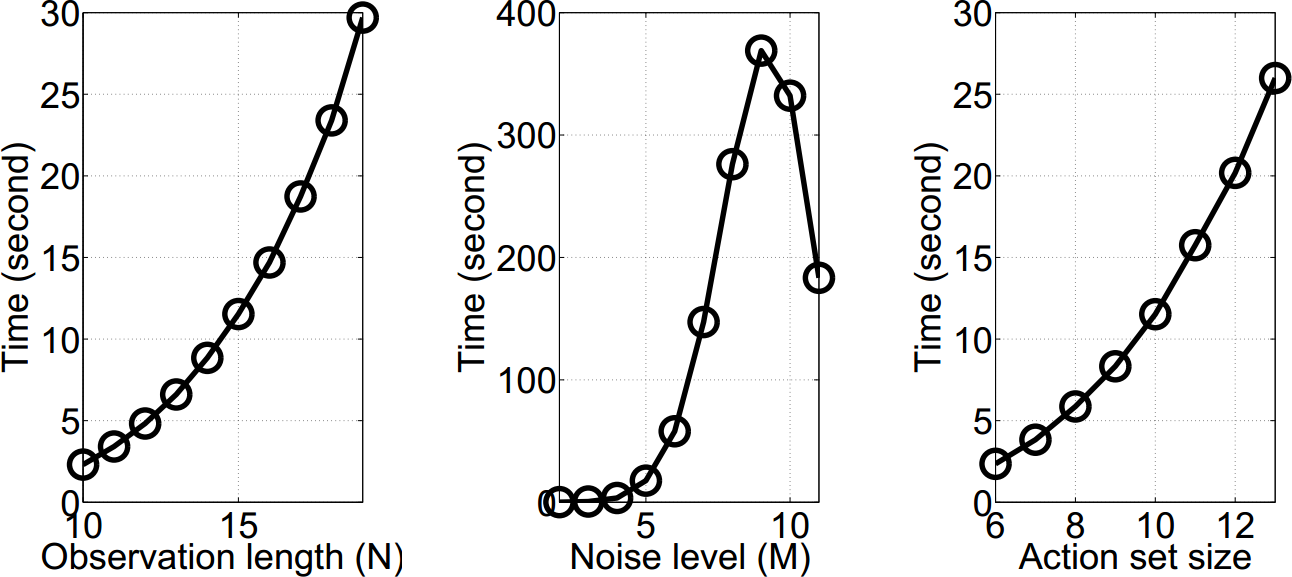}
\caption{Brute force time cost for calculating $P(\mathbf Y| C)$}
\label{fig:DPTime}
\end{figure}

\begin{figure}[!ht]
\centering
\includegraphics[width=0.85\columnwidth] {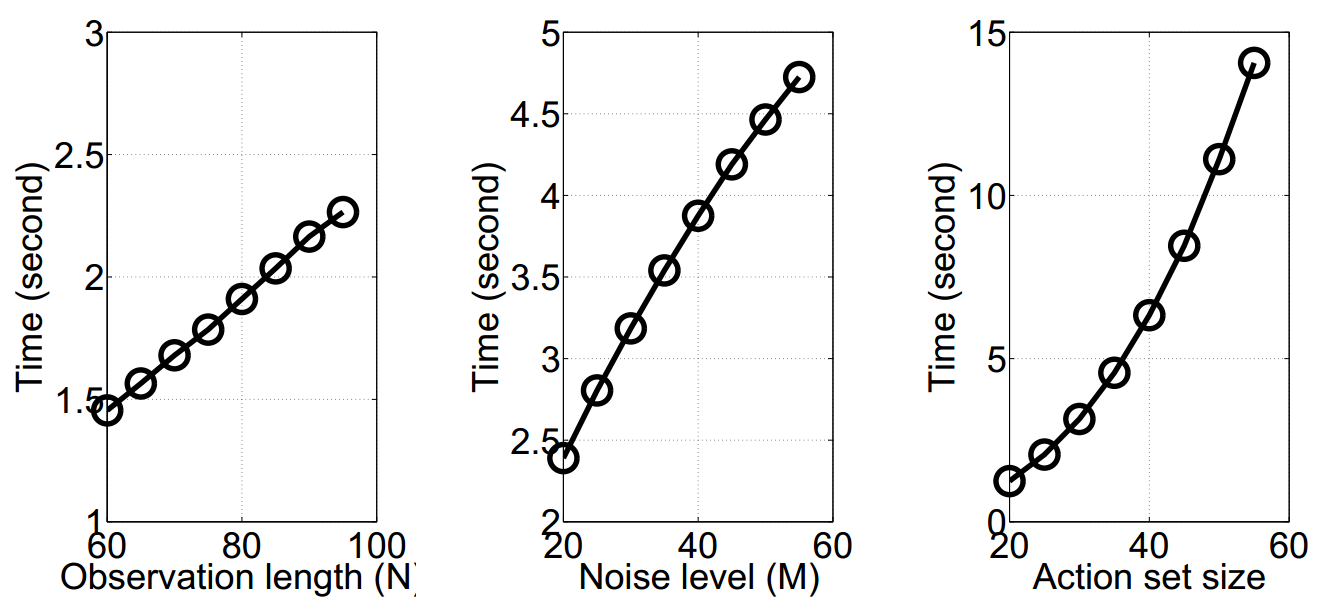}
\caption{Proposed algorithm time cost for calculating $P(\mathbf Y| C)$}
\label{fig:DPTime2}
\end{figure}

The time cost\footnote{The prototype system was written in Python and was executed on a desktop with i5 CPU and 16G RAM.} comparison between approximation and exact calculation is given in Figures \ref{fig:DPTime} and \ref{fig:DPTime2}. The exact calculation time cost increases exponentially as the observation length $N$ increases. On the other hand, the time costs via Monte-Carlo approximation have relatively small variations (between 1 to 2.5 seconds) for different observation lengths. Similar phenomena can be observed as the noise level $M$ or the action set size $\Omega$ increases. The reason is that no matter how many possible values $\mathbf X$ can take, only a fixed number of samples ($30,000$ in our case) are taken to estimate the ECA. To verify the noise inference algorithm, we use brute-force algorithm to calculate $P(\mathbf Y=\mathbf Y_0)$ by summing over $\mathbf X$ that satisfies the constraints. In comparison, we use dynamic programming to calculate $P(\mathbf Y=Y_0)$. In summary, the time cost with Monte-Carlo sampling is within a reasonable range for Algorithm \ref{alg:approximation}.

\section{Impact Analysis of Obfuscation Models}
\label{sec:simulation}
\subsection{Simulation Design}
\label{sec:simulation_setup}
\begin{figure*}[!ht]
\centering
\includegraphics[width=1.8\columnwidth]{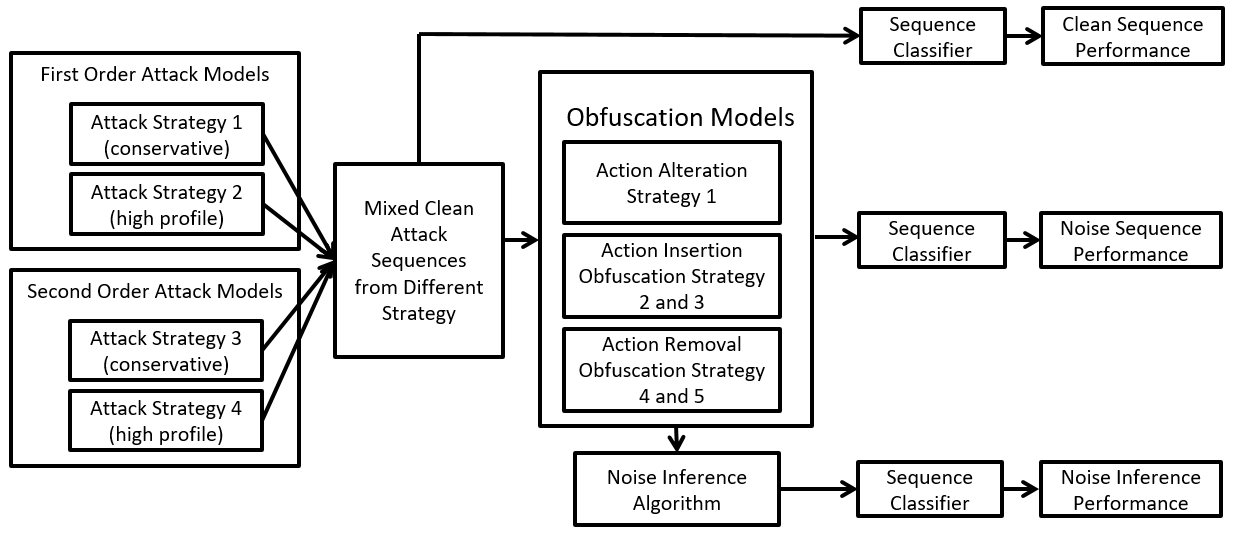}
\caption{Simulation overview for attack sequence classification with obfuscations}
\label{fig:ch5_sim_overview}
\end{figure*}

The simulation setup is shown in Fig. \ref{fig:ch5_sim_overview}. The experiment considers 4 attack models and 5 obfuscation models. The clean attack sequences are generated from the attack models. Mixed attack sequences from different models feed into different attack obfuscation techniques. The sequence classifiers for clean sequences and noise sequences assume the full knowledge of the attack model but not the attack obfuscation model. The likelihoods are compared for a given observed sequence across different generative models to perform classification. On the other hand, the noise inference algorithm utilizes the knowledge of the attack obfuscation models to calculate the distribution of the obfuscated sequence to make the sequence classifier better. We compare the classification results for the clean attack sequence and the noise attack sequence using the proposed inference algorithm. The ECA defined in Section \ref{sec:sec3problem_statement} is the primary performance metric. 

\begin{figure*}[!ht]
\centering
\includegraphics[width=1.8\columnwidth]{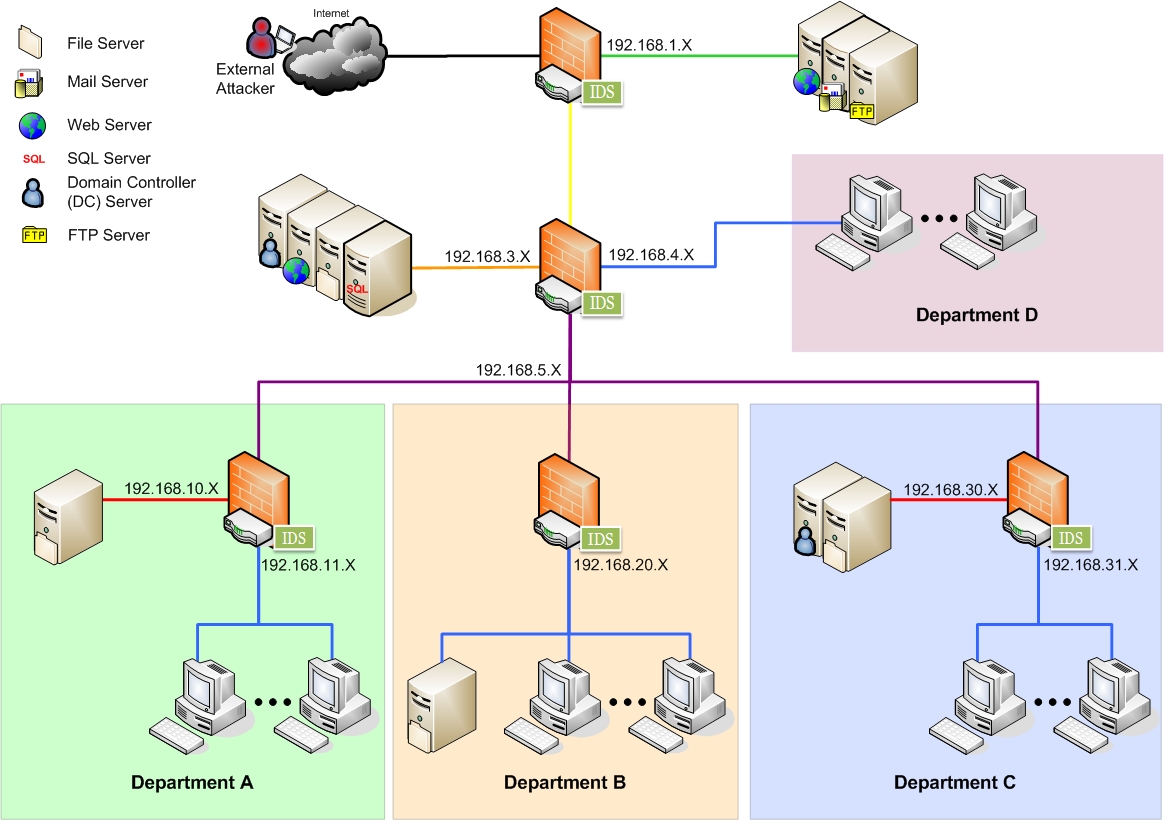}
\caption{Network used for simulation}
\label{fig:ch5_bvt}
\end{figure*}

Figure \ref{fig:ch5_bvt} gives the reference network diagram used for the experiment. It describes a small enterprise network with six subnets, eleven servers and four clusters of hosts (24 hosts in total). The whole network has 31 open services (15 types total) and interconnected via four routers. The attacker began the attack from the Internet. The external servers (web server and file server) were first attacked with \textit{abuse of functionality}. After a few steps, the attacker obtained the vulnerability information and performed a buffer overflow attack on the file server and compromised the external file server 192.168.1.3. Using this stepping stone, the attacker compromised the internal server (Domain controller 192.168.3.1) and use it to probe the hosts in Department C (192.168.30.x). 

In general, one can model network attacks at various levels with a combination of attributes reported by NIDS and host logs. To demonstrate the use of the proposed framework, this paper considers $15$ classical, widely used attack actions from five categories selected from MITRE's common attack pattern enumeration and classification \cite{capec}. The attack categories and action space $\Omega$ are shown in Fig. \ref{fig:actionspace}. 

\begin{figure}[!ht]
\centering
\begin{itemize}
\item Abuse of functionality (C-0)
\begin{itemize}
\item Detect unpublicised web pages and services (A-0-1)
\item Directory traversal (A-0-2)
\item Web server application fingerprinting (A-0-3)
\end{itemize}
\end{itemize}
\begin{itemize}
\item Network reconnaissance (C-1)
\begin{itemize}
\item Infrastructure-based footprinting (A-1-1)
\item Host discovery (A-1-2)
\item Scanning for vulnerable software (A-1-3)
\end{itemize}
\end{itemize}
\begin{itemize}
\item Probabilistic techniques (C-2)
\begin{itemize}
\item Fuzzing (A-2-1)
\item Screen temporary files for information (A-2-2)
\item Client-server protocol manipulation (A-2-3)
\item Dictionary-based password attack (A-2-4)
\end{itemize}
\end{itemize}
\begin{itemize}
\item Buffer overflow and code injection (C-3)
\begin{itemize}
\item Manipulating user-controlled variables (A-3-1)
\item Command or script injection (A-3-2)
\item Hijacking a privileged process / thread (A-3-3)
\end{itemize}
\item Data leakage attacks (C-4)
\begin{itemize}
\item Data excavation attacks (A-4-1)
\item Data interception / sniffer attacks (A-4-2)
\end{itemize}
\end{itemize}
\caption{An example of action space (attack patterns)}
\label{fig:actionspace}
\end{figure}

The five categories (C-0 to C-4) show different levels (stages) of the intrusion process. Abuse of functionality is a low-profile information-gathering step; taking advantage of the function provided by the target system can achieve a certain level of information collection without leaving much malicious trace, because such functions are designed to serve normal requests. For example, instead of scanning the target web server to get the server version, one can try to access a non-existing web page and observe the HTTP-404-ERROR generated by the server, which can expose the server platform and version. Network reconnaissance is a category of high-profile scanning in addition to abuse of functionality. Actions in this category are essentially taking advantage of the TCP/IP protocol, \eg, TRACEROUTE, PING, NMAP, \etc, to explore unknown environment. Probabilistic techniques represent another type of exploration, using a number of trials to identify vulnerabilities. For example, fuzzing is widely used in software testing by feeding the system with invalid, unexpected random inputs. By observing the system feedback, an experienced attacker can discover possible design flaws, including the chance of getting buffer overflow or code injection vulnerability, which is part of category C-3 and can eventually compromise the target machine. After compromising the target the ultimate goal of an attack can be stealing sensitive data or data excavation, \eg, generic cross-browser cross-domain thefts \cite{capec}, or data interception/sniffer.

We consider four attack models, two first-order (Strategy-1 and Strategy-2) and two second-order (Strategy-3 and Strategy-4) ones. As pointed out by Fava \etal \cite{Fava2008} and Du \etal \cite{Du2010}, most attack behaviors can be captured with first and second-order Markov models. Furthermore, higher order models can be too specific with high complexity and perform poorly because of Bias-variance trade-off \cite{hastie2009}. The two models are inspired from real attacks in ICTF hacking competition data set \cite{ICTFData} \cite{Childers2010} and CAIDA data set \cite{CAIDAData} \cite{Moore2004}. Attack Strategy-1 can be explained as a \textit{two phases attack}: reconnaissance and intrusion. The attacker is more hesitant to switch between phases than stay within a phase. The specific probability numbers in the table reflect the characteristics of the attack, \eg, the automatic script the attacker is using. In fact, our experience suggests that the probabilities of action transitions are quite reliable for detecting the attack tools such as Metasploit \cite{Maynor2007} or Nessus \cite{Beale2004a}. Attack Strategy-2 shows a different attack strategy: the attacker utilizes the reconnaissance actions throughout the attack process, and perform specific exploits only sporadically. Attack Strategy-3 and Attack Strategy-4 are much more complicated models since the second-order Markov transition matrix has $15$ times more parameters (the size of action space). The overall idea for Attack Strategy-3 and Attack Strategy-4 is very similar to Attack Strategy-1 and Attack Strategy-2, respectively, but more parameters are used to describe the specific attack behavior. For example, Attack Strategy-3 describes the attacker spending more time on C-1 reconnaissance stage and we have higher chance to observe a long sequence of A-0-1. Attack Strategy-4 is a high-profile attack but with specific long sequences of certain vulnerability attempts. 

For action alternation obfuscation model, the Obfuscation Strategy-1 is mostly trying to exchange actions in the same category, such as Abuse of functionality, Network reconnaissance, etc. For action insertion obfuscation, two specific scenarios are designed for simulation. Obfuscation Strategy-2, describes an attacker injecting \textit{independent noise observations}, which means that the injected actions have nothing to do with the previous attack action and the "clean attack actions". The injected actions have its own distribution conditionally independent to other random variables. For example, the injected noisy actions can have 80\% of \textit {abuse of service} action, and 20\% of \textit{network reconnaissance} action. Obfuscation Strategy-3 describes a more complicated action injection plan, the injected actions depend on the clean actions. For example, in the network reconnaissance stage and vulnerability attempt stage, the attacker would have different preference to inject more actions in some categories than other categories. Such noise injection plan can be effective to confuse the alert analysis engine on the \textit{intrusion stage assessment} and conceal the intrusion stage and the real intent.  

For action removal obfuscation, Obfuscation Strategy-4 describes the attacker attempting on different services and vulnerabilities with a mixture of attack actions such as buffer overflow and abuse of functionality over different services. Note that, depending on the configuration, an NIDS may detect some but not all different actions. Obfuscation Strategy-5 represents more stealthy and decoy attacks. There are covering-up actions with intended actions. We assume the covering-up actions conditionally depend on the intended actions. This is different from executing random actions because the goal of such obfuscation is to mislead the analyst, the noise sequence $\mathbf Y$ is linked to the previous noisy actions.

\subsection{The Impact of Alteration, Insertion, vs. Removal}

The algorithm proposed in this paper enables us to assess the impact of attack obfuscations. We first evaluate the ECA with the three types of obfuscation models (Alteration, Insertion vs. Removal). The algorithms proposed in this paper calculate $P(\mathbf Y)$ efficiently, and $P(\mathbf X)$ can be directly derived based on the given attack models. Note that $\max_C P(C|\mathbf Y)$ represents the best likelihood one can match a given sequence $\mathbf Y$ to a model $C$, and the $1-\max_C P(C|\mathbf Y)$ also implies the least error for the given sequence and the models. To assess the overall impact to any sequence, clean or obfuscated, that can occur under the attack models, one will need to calculate $\sum_\mathbf X P(\mathbf X)\max_C P(C|\mathbf X)$ and $\sum_\mathbf Y P(\mathbf Y)\max_C P(C|\mathbf Y)$. 

The simulation results shown below can be interpreted by comparing the ``InfAlg'' case with the ``noise'' case. The ``InfAlg'' case gives the ECA when the proposed algorithms are used to infer/classify the obfuscated attack sequences to the original attack model, whereas the ``noise'' case is when the obfuscated sequences are directly classified to the best matched attack model without using the inference algorithms. The improvement in ECA from the ``noise'' case to the ``InfAlg'' case gives the recovery or improvement in classification accuracy using the proposed algorithms. In addition, we present the ``clean'' case to reflect the ideal scenario where classification is performed on attack sequences where no obfuscation is done, which is the best one can ever recover for ECA.

\begin{figure*}[!ht]
\centering
\subfigure[Impact for action insertion models and action removal models]{
\includegraphics[width=0.98\columnwidth]{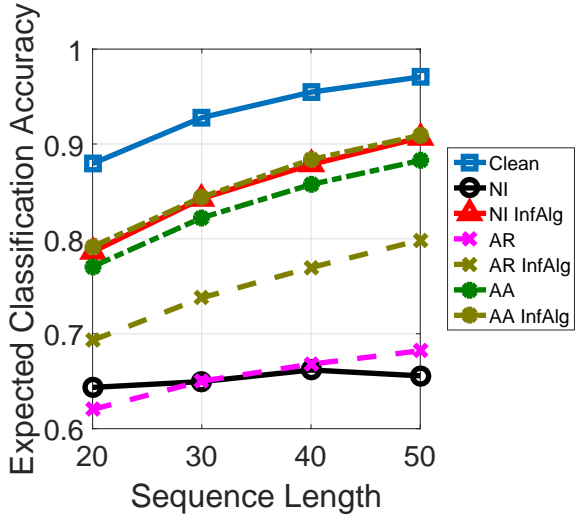}
\label{fig:ch5_sim_res_insertion}
}
\centering
\subfigure[Impact for first and second-order attack models]{
\includegraphics[width=0.88\columnwidth]{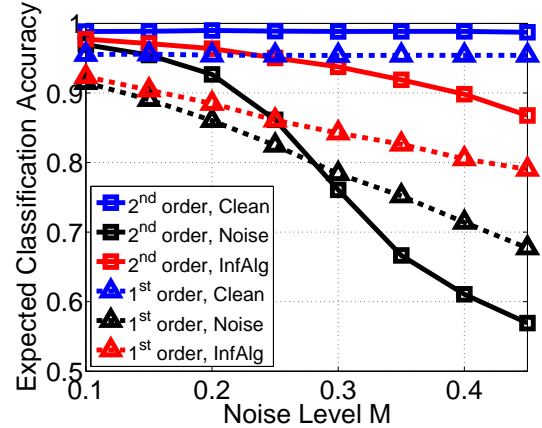}
\label{fig:s_vs_N}}
\caption{ECA for Noise Insertion (NI), Action Removal (AR) and Action Alteration (AA) obfuscations and comparison between the first and second-order attack models}
\label{fig:ch5_overall_impact}
\end{figure*}

Figure \ref{fig:ch5_sim_res_insertion} shows the inference algorithm achieves more effective ECA improvements for Noise Insertion (NI) and Action Removal (AR) with approximately 20\%, comparing to the 5\% in the Action Alteration (AA) case. It also shows that the ECA increases as the observation length increases, \eg, the more observations of the same attack behavior, the easier one can classify sequences to the correct attack models. Figure \ref{fig:ch5_overall_impact} compares the ECA when the first and second-order attack models are used. We observed that when the obfuscation level increases, the performance drops, especially for the second-order model case without inference. At around the obfuscation level of 28\%, the obfuscated sequences without inference for the second-order case actually begin to exhibit worse performance than that for the first-order case. Fortunately, even with the limited knowledge of obfuscation, the optimal classification rate can be recovered, \eg, from 60\% to 90\% when the obfuscation level is at 40\%. Interestingly, the performance recovered through inference for the second-order model case is better than that for the first-order case, at least up to the 45\% obfuscation level, which is very high. Generally speaking, the higher the obfuscation level, the more improvements one can achieve, for both first and second-order cases. The performance recovered through inference is closer to the absolute limit exhibited by the clean curves for the second-order model case than that for the first-order model case, at least when the obfuscation level is not too high.

\subsection{The impact of obfuscation level estimation}
\label{sec:sim_res_inaccurate_noise_model}

One important parameter in obfuscation model for action alteration is the obfuscation level, \eg, how much action alteration exists in the sequence. In order to run the noise inference algorithm, this parameter $M$ is assumed to be known. In real situation, one may argue that it is not reasonable for security analysts to know how an attacker performs obfuscation in such a detailed level. Therefore, in this subsection, we want to investigate how the parameter $M$ impact the proposed algorithm, and how much impact the inaccurate estimation of $M$ can cause empirically. We will show that, only an approximation of the noise level is needed to get a reasonably good inference results. In other words, as long as the analysts have a rough estimation of the level of obfuscation, the proposed algorithm will be useful to classify attack sequences. In our simulation, the obfuscated attack sequences were created using the true obfuscation level value $M_{true}$, while the estimated obfuscation level $M_{est}$ is intentionally set to deviate from $M_{true}$. The algorithm is executed based on the $M_{est}$ value, to assess how ECA might be different when the estimated value is not accurate. 

\begin{figure*}
\centering
\subfigure[Inaccurate M estimation with respect to different sequence length]{
\centering
\includegraphics[width=0.95\columnwidth]{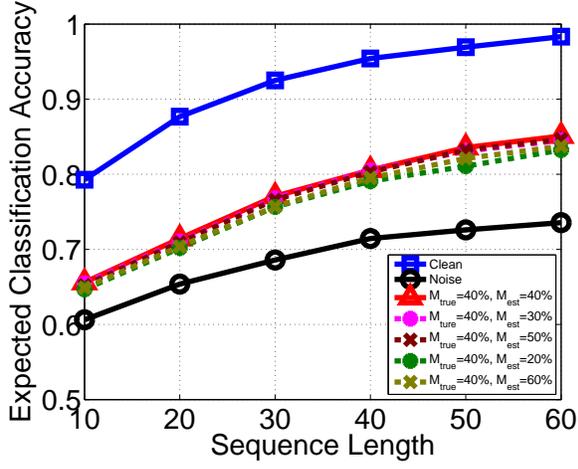}
\label{fig:InAccM1}
}
\subfigure[Inaccurate M estimation with respect to different noise level]{
\centering
\includegraphics[width=0.95\columnwidth]{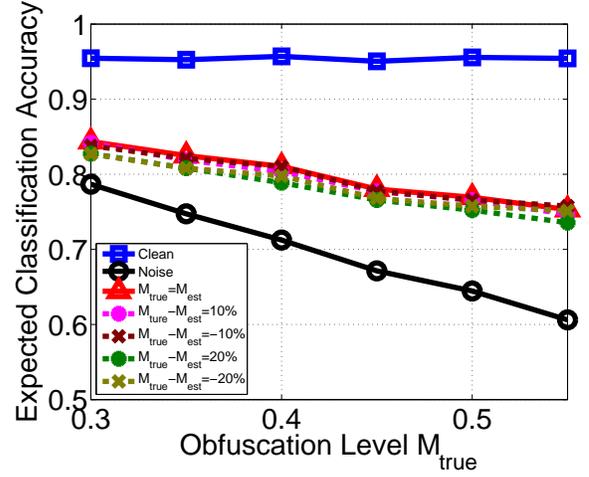}
\label{fig:InAccM2}
}
\subfigure[Inaccurate M estimation with respect to different noise level]{
\centering
\includegraphics[width=0.95\columnwidth]{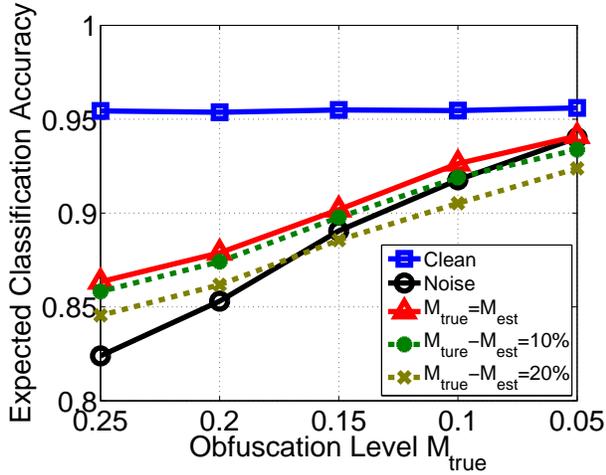}
\label{fig:m_app_0}
}
\subfigure[Inaccurate M estimation with respect to different noise level]{
\centering
\includegraphics[width=0.95\columnwidth]{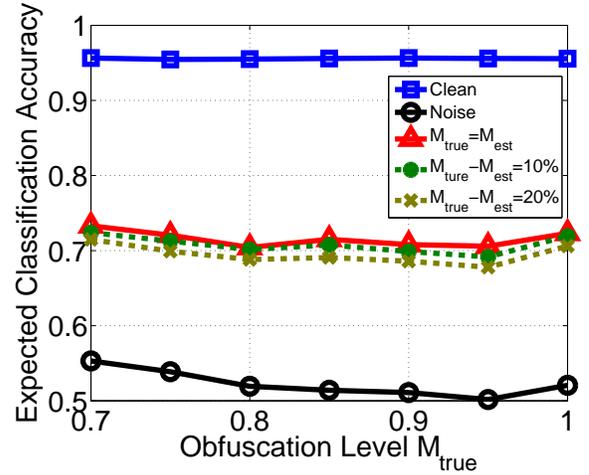}
\label{fig:m_app_1}
}
\caption{Action Alteration Obfuscation, Inaccurate Noise Model Estimation Impact Simulation}
\label{fig:m_discussion}
\end{figure*}

Figure \ref{fig:m_discussion} shows the ECA under various obfuscation level estimations. Figure \ref{fig:InAccM1} shows the inaccurate $M$ when the observation length $N$ ranges from 10 to 60. The real obfuscation level $M_{true}$ is 40\%. The estimated $M_{est}$ ranges from 20\% to 60\%. Obviously, ECA is the highest when $M_{true}=M_{est}=40\%$. Also, the closer $M_{est}$ is to $M_{true}$, the better performance it has. Interestingly, the ECA does not change much as $M_{est}$ moves away from $M_{ture}$, this is true even for different observation lengths.

Figure \ref{fig:InAccM2} evaluates the effect of $M_{est}$ as $M_{true}$ increases from 30\% to 55\%. We test the cases when $|M_{est}-M_{true}|=10\%$ and 20\% with a fixed observation length of 40. The results are similar to that from Fig. \ref{fig:InAccM1}. The larger the difference between $M_{est}$ and $M_{true}$ the lower the ECA. However the impact of this difference is insignificant. Again these observations remain the same as the noise level increases from 30\% to 55\%.

Figure \ref{fig:m_app_0} and \ref{fig:m_app_1} give the ECA when noise level $M$ approaches two extreme cases, \eg, $M=0\%$ and $M=100\%$. Note that, for $M=0\%$ the inference algorithm will not work, because $P(\mathbf Y| \mathbf X)=0$. At the extreme low noise level, inaccurate estimation of $M$ will actually have slightly worse ECA when using the inference algorithm.

As $M$ approach $100\%$, the results are shown in Figure \ref{fig:m_app_1}. In this case, the inference algorithm can always make improvements and the inaccurate $M$ estimation does not affect the ECA too much. Furthermore, ECA value remains similar for $M_{true}$ ranging from 70\% to 100\%. This can be explained with the combinatorial number of possible noise sequences. Specifically, as discussed earlier, the possible noise sequences for a given $M$ is $N \choose M$ $(|\Omega|-1)^M$. Therefore, the number of possible sequences will reduce when $M=100\%$, \eg, the number of changes equals to the sequence length. A similar trend was observed in the time cost evaluation results (Fig. \ref{fig:DPTime})

\section{Conclusion}
\label{sec:conclusion}
The mixture of organized cyber crimes and random attacks against enterprise and government networks has led to asymmetric cyber battlefields filled with large-scale cyber attacks. To obtain a timely situation awareness from overwhelming and noisy observations, defense can benefit from probabilistic graphical models to inference true attacks from obfuscated observations.

This work developed a general framework to model attack alteration, insertion, and removal obfuscation strategies. To evaluate the impact for specific attack scenarios, The ECA metric has been proposed, which enables the study of the benefits and limitations of attack sequence modeling and classification. Comprehensive simulations on various combination of attack models and obfuscation techniques show that the noise insertion has the highest impact, \ie, gives the lowest ECA when breaking the attack sequence pattern by random noise actions. Fortunately, the inference algorithm proposed can recover ECA well in most cases. In general, our experimental results show that as the observed sequence length increases, or as the obfuscation level decreases, the ECA can improve. As the attack model complexity increases, attack obfuscation can cause higher impact, but the inference algorithm can also recover more. The proposed inference algorithm is also shown to be robust when the estimated level of obfuscation is not accurate.

The attack obfuscation modeling framework and the inference algorithms developed here can be applied to other contexts beyond analyzing network attacks. Any observed sequences that might suffer from noise and require matching to pre-defined models can use this work to recover the most likely original model or evaluate quantitatively the optimal performance one can achieve to separate the observed instances.

\ifCLASSOPTIONcompsoc
  \section*{Acknowledgments}
\else
  \section*{Acknowledgment}
\fi
This research is partial supported by National Science Foundation Award \#1526383.
\ifCLASSOPTIONcaptionsoff
  \newpage
\fi

\bibliographystyle{IEEETran}
\bibliography{ref}

\begin{thebibliography}{10}
\providecommand{\url}[1]{#1}
\csname url@samestyle\endcsname
\providecommand{\newblock}{\relax}
\providecommand{\bibinfo}[2]{#2}
\providecommand{\BIBentrySTDinterwordspacing}{\spaceskip=0pt\relax}
\providecommand{\BIBentryALTinterwordstretchfactor}{4}
\providecommand{\BIBentryALTinterwordspacing}{\spaceskip=\fontdimen2\font plus
\BIBentryALTinterwordstretchfactor\fontdimen3\font minus
  \fontdimen4\font\relax}
\providecommand{\BIBforeignlanguage}[2]{{%
\expandafter\ifx\csname l@#1\endcsname\relax
\typeout{** WARNING: IEEEtran.bst: No hyphenation pattern has been}%
\typeout{** loaded for the language `#1'. Using the pattern for}%
\typeout{** the default language instead.}%
\else
\language=\csname l@#1\endcsname
\fi
#2}}
\providecommand{\BIBdecl}{\relax}
\BIBdecl

\bibitem{Ptacek1998}
T.~H. Ptacek and T.~N. Newsham, ``Insertion, evasion, and denial of service:
  Eluding network intrusion detection,'' DTIC Document, Tech. Rep., 1998.

\bibitem{Lyon2009a}
G.~Lyon, \emph{Nmap Network Scanning: The Official Nmap Project Guide to
  Network Discovery and Security Scanning}.\hskip 1em plus 0.5em minus
  0.4em\relax Insecure Publishing, 2009.

\bibitem{Fuchsberger2005}
A.~Fuchsberger, ``{Intrusion Detection Systems and Intrusion Prevention
  Systems},'' \emph{Information Security Technical Report}, vol.~10, no.~3, pp.
  134--139, 2005.

\bibitem{Du2014}
H.~Du and S.~Yang, ``{Probabilistic Inference for Obfuscated Network Attack
  Sequences},'' in \emph{Proceedings of the 44th Annual IEEE/IFIP International
  Conference on Dependable Systems and Networks (DSN 2014)}.\hskip 1em plus
  0.5em minus 0.4em\relax IEEE, 2014, pp. 1--11.

\bibitem{Du2010}
H.~Du, D.~Liu, J.~Holsopple, and S.~Yang, ``{Toward Ensemble Characterization
  and Projection of Multistage Cyber Attacks},'' in \emph{Proceedings of the
  19th IEEE International Conference on Computer Communications and Networks
  (ICCCN'10)}, 2010, pp. 1--8.

\bibitem{Wang2006}
L.~Wang, A.~Liu, and S.~Jajodia, ``Using attack graphs for correlating,
  hypothesizing, and predicting intrusion alerts,'' \emph{Computer
  Communications}, vol.~29, no.~15, pp. 2917 -- 2933, 2006.

\bibitem{Du2011f}
H.~Du and S.~Yang, ``{Discovering collaborative cyber attack patterns using
  social network analysis},'' in \emph{Proceedings of Social Computing,
  Behavioral-Cultural Modeling and Prediction (SBP'10)}.\hskip 1em plus 0.5em
  minus 0.4em\relax Springer, 2011, pp. 129--136.

\bibitem{Yang2014}
S.~J. Yang, H.~Du, J.~Holsopple, and M.~Sudit, \emph{Attack Projection}.\hskip
  1em plus 0.5em minus 0.4em\relax Springer, 2014.

\bibitem{DARPACINDER}
\BIBentryALTinterwordspacing
{Defense Advanced Research Projects Agency (DARPA) Cyber Insider Threat
  Program}. (Access Date: Oct. 2017). [Online]. Available:
  \url{https://www.fbo.gov/index?s=opportunity&mode=form&tab=core&id=585e02a51f77af5cb3c9e06b9cc82c48}
\BIBentrySTDinterwordspacing

\bibitem{Haines2003}
J.~Haines, D.~Ryder, L.~Tinnel, S.~Taylor, and D.~{Kewley Ryder}, ``{Validation
  of sensor alert correlators},'' \emph{IEEE Security \& Privacy}, vol.~1,
  no.~1, pp. 46--56, Jan. 2003.

\bibitem{rabiner1989}
L.~R. Rabiner, ``A tutorial on hidden markov models and selected applications
  in speech recognition,'' \emph{Proceedings of the IEEE}, vol.~77, no.~2, pp.
  257--286, 1989.

\bibitem{Du2011b}
H.~Du and S.~J. Yang, ``{Characterizing transition behaviors in Internet attack
  sequences},'' in \emph{Proceedings of the 20th IEEE International Conference
  on Computer Communications and Networks (ICCCN'11)}, 2011, pp. 1--6.

\bibitem{Du2013a}
H.~Du and S.~Yang, ``Temporal and spatial analyses for large-scale cyber
  attacks,'' \emph{Handbook of Computational Approaches to Counterterrorism},
  vol.~1, no.~2, pp. 559--578, 2013.

\bibitem{murphy2002}
K.~P. Murphy, ``Dynamic bayesian networks: representation, inference and
  learning,'' Ph.D. dissertation, University of California, 2002.

\bibitem{hastie2009}
T.~Hastie, R.~Tibshirani \emph{et~al.}, \emph{The Elements of Statistical
  Learning: Data Mining, Inference and Prediction}, 2nd~ed.\hskip 1em plus
  0.5em minus 0.4em\relax Springer, 2009.

\bibitem{Bishop2006}
C.~M. Bishop, \emph{Pattern Recognition and Machine Learning (Information
  Science and Statistics)}.\hskip 1em plus 0.5em minus 0.4em\relax Secaucus,
  NJ, USA: Springer-Verlag New York, Inc., 2006.

\bibitem{serfling1974}
R.~Serfling, ``Probability inequalities for the sum in sampling without
  replacement,'' \emph{The Annals of Statistics}, vol.~2, no.~1, pp. 39--48,
  1974.

\bibitem{capec}
\BIBentryALTinterwordspacing
{Common Attack Pattern Enumeration and Classification. \url{URL:
  http://capec.mitre.org}, Access Date: Aug. 2013}. [Online]. Available:
  \url{http://capec.mitre.org}
\BIBentrySTDinterwordspacing

\bibitem{Fava2008}
D.~Fava, S.~Byers, and S.~Yang, ``Projecting cyberattacks through
  variable-length markov models,'' \emph{IEEE Transactions on Information
  Forensics and Security}, vol.~3, no.~3, pp. 359--369, Sept. 2008.

\bibitem{ICTFData}
\BIBentryALTinterwordspacing
{UCSB International Capture The Flag (ICTF) Hacking Competition Data Set}.
  (Access Date: Oct. 2017). [Online]. Available: \url{http://ictf.cs.ucsb.edu/}
\BIBentrySTDinterwordspacing

\bibitem{Childers2010}
N.~Childers, B.~Boe, L.~Cavallaro, L.~Cavedon, M.~Cova, M.~Egele, and G.~Vigna,
  ``{Organizing Large Scale Hacking Competitions},'' in \emph{Detection of
  Intrusions and Malware, and Vulnerability Assessment (DIMVA)}, vol.
  6201.\hskip 1em plus 0.5em minus 0.4em\relax Springer, 2010, pp. 132--152.

\bibitem{CAIDAData}
\BIBentryALTinterwordspacing
{The CAIDA UCSD network telescope two days in November 2008 dataset}. (Access
  Date: Oct. 2017). [Online]. Available:
  \url{http://www.caida.org/data/overview/}
\BIBentrySTDinterwordspacing

\bibitem{Moore2004}
D.~Moore, C.~Shannon, G.~Voelker, and S.~Savage, ``{Network telescopes:
  technical report},'' Tech. Rep., 2004.

\bibitem{Maynor2007}
D.~Maynor and K.~Mookhey, \emph{Metasploit toolkit for penetration testing,
  exploit development, and vulnerability research}.\hskip 1em plus 0.5em minus
  0.4em\relax Syngress Publishing, 2007.

\bibitem{Beale2004a}
J.~Beale \emph{et~al.}, \emph{Nessus network auditing}.\hskip 1em plus 0.5em
  minus 0.4em\relax Syngress Publishing, 2004.

\end{thebibliography}

\begin{IEEEbiography}[{\includegraphics[width=1in,height=1.25in,clip,keepaspectratio]{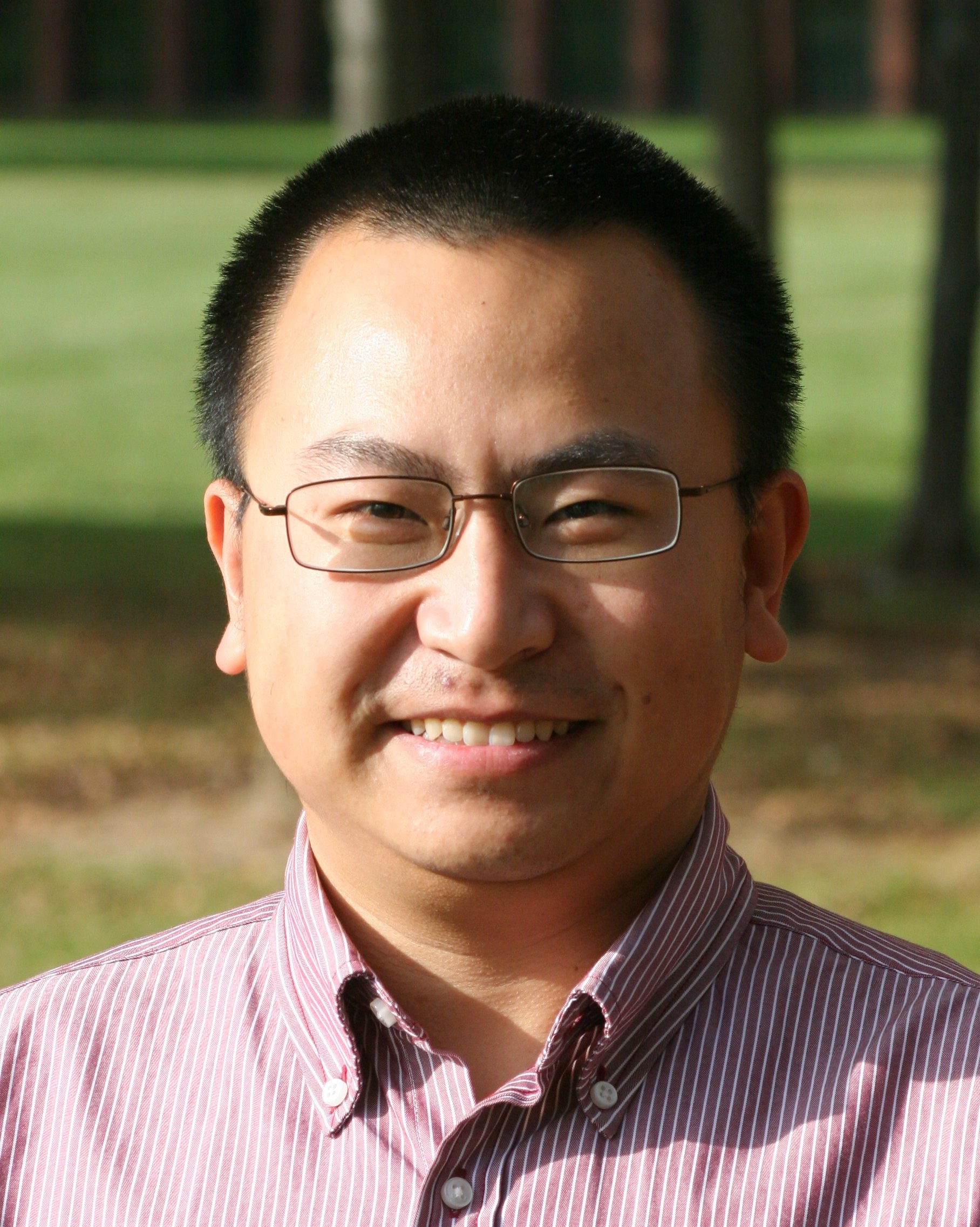}}]{Haitao Du}
Dr. Haitao Du received his B.S. degree in Telecommunications Engineering from Xidian University, Xi'an, China in 2006, and his Ph.D. degree in Computing and information sciences from Rochester Institute of technology, 2014. He is currently a data scientist at K12 inc. and driving the process building big data solutions that achieve predictive modeling on student academic performance and retention.
\end{IEEEbiography}
\begin{IEEEbiography}[{\includegraphics[width=1in,height=1.25in,clip,keepaspectratio]{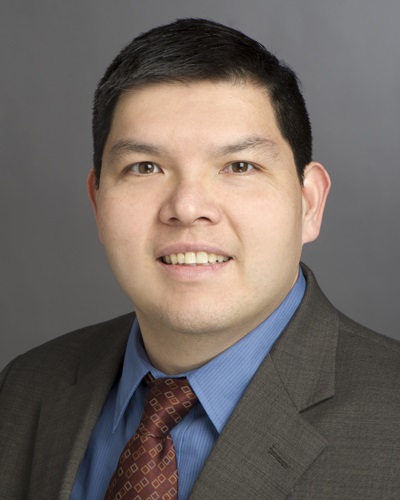}}]{Shanchieh Jay Yang}
Dr. S. Jay Yang received his B.S. degree in Electronics Engineering from National Chiao-Tung University, Hsin-Chu, Taiwan in 1995, and his M.S. and Ph.D. degrees in Electrical and Computer Engineering from the University of Texas at Austin in, 1998 and 2001, respectively. He is currently a Professor and the Department Head for the Department of Computer Engineering at RIT. He and his research group has developed several systems and frameworks in the area of cyber attack modeling for predictive situation, threat and impact assessment. He has published more than sixty papers and was invited as a keynote speaker, a panelist, and a guest speaker in various venues. He was a co-chair for IEEE Joint Communications and Aerospace Chapter in Rochester NY in 2005, when the chapter was recognized as an Outstanding Chapter of Region 1. He has also contributed to the development of two Ph.D. programs at RIT, and received Norman A. Miles Award for Academic Excellence in Teaching in 2007.
\end{IEEEbiography}

\end{document}